\documentclass[twocolumn,trackchanges]{aastex62}
\usepackage{booktabs}
\usepackage[flushleft]{threeparttable}

\def\code#1{\texttt{#1}}

\received{January 2, 2020}
\revised{March 3, 2020}
\accepted{March 30, 2020}

\submitjournal{ApJ}

\shorttitle{Gravitational Wave Treasure map }
\shortauthors{Wyatt et al.}

\begin{document}

\title{The Gravitational Wave Treasure Map: A Tool to Coordinate, Visualize, and Assess the Electromagnetic Follow-Up of Gravitational Wave Events}

\correspondingauthor{Samuel D. Wyatt}
\email{swyatt@email.arizona.edu}

\author[0000-0003-2732-4956]{Samuel D. Wyatt}
\affiliation{Steward Observatory, University of Arizona, 933 North Cherry Avenue, Rm. N204, Tucson, AZ 85721-0065, USA}

\author[0000-0002-2810-8764]{Aaron Tohuvavohu}
\affiliation{Department of Astronomy and Astrophysics, University of Toronto, Toronto, ON, CA}

\author[0000-0001-7090-4898]{Iair Arcavi}
\affiliation{The School of Physics and Astronomy, Tel Aviv University, Tel Aviv 69978, Israel}
\affiliation{CIFAR Azrieli Global Scholars program, CIFAR, Toronto, Canada}

\author[0000-0001-9589-3793]{Michael J. Lundquist}
\affiliation{Steward Observatory, University of Arizona, 933 North Cherry Avenue, Rm. N204, Tucson, AZ 85721-0065, USA}

\author[0000-0003-4253-656X]{D. Andrew Howell}
\affiliation{Department of Physics, University of California, Santa Barbara, CA 93106-9530}
\affiliation{Las Cumbres Observatory, 6740 Cortona Drive, Suite 102, Goleta, CA 93117-5575, USA}

\author[0000-0003-4102-380X]{David J. Sand}
\affiliation{Steward Observatory, University of Arizona, 933 North Cherry Avenue, Rm. N204, Tucson, AZ 85721-0065, USA}

\begin{abstract}

We present the Gravitational Wave Treasure Map, a tool to coordinate, visualize, and assess the electromagnetic follow-up of gravitational wave (GW) events.  With typical GW localization regions of hundreds to thousands of square degrees and dozens of active follow-up groups, the pursuit of electromagnetic (EM) counterparts is a challenging endeavor, but the scientific payoff for early discovery of any counterpart is clear. With this tool, we provide a website and API interface that allows users to easily see where other groups have searched and better inform their own follow-up search efforts. A strong community of Treasure Map users will increase the overall efficiency of EM counterpart searches and will play a fundamental role in the future of multi-messenger astronomy.

\end{abstract}

 
\section{Introduction}\label{sec:intro}

The era of gravitational wave multi-messenger astronomy has begun with the ground-breaking discoveries of the LIGO-Virgo Collaboration (LVC) and their network of gravitational wave detectors.   In their first two observing runs, the  advanced LIGO and Virgo observatories discovered 11 gravitational wave (GW) events including ten binary black hole mergers and one binary neutron star merger \citep{lvc_O1O2}. The third observing run (O3) began on 1 April 2019 and has greatly increased the number of likely GW events with detections of numerous binary black hole mergers (e.g. \citealt{GCN24069,GCN24098,GCN24141})  and several mergers which likely involved neutron stars (e.g. \citealt{GCN24168,GCN24237,GCN24462}).

Of the numerous GW events that have been detected by the Advanced LIGO and Virgo observatories, only one, GW170817, has had an identified electromagnetic (EM) counterpart \citep{GW170817MMA,Arcavi17_2,Coulter17,Lipunov17,Tanvir17,Soares17,Valenti2017, evans2017,Fermi_KN,INTEGRAL_KN,Haggard_17,Troja_17,Hallinan_17,Margutti_17}. As we are learning during the course of O3, GW170817 was atypical in several respects.  First, GW170817 was very nearby, at a distance of $\approx$40 Mpc \citep{gcn21513}.  Second, it was very well localized, ultimately to $\approx$28 deg$^2$ \citep{lvc_gw170817}.   Third, it had a coincident gamma-ray detection from \emph{Fermi/GBM} and \emph{INTEGRAL/SPI-ACS} \citep{gw170817_grb}. During O3, identifying these EM counterparts has been an observational challenge with typical GW events at a few hundred Mpc and localized to $\sim$$10^{2-3}$~deg$^2$ on the sky.  Despite the challenge, early identification of counterparts is critical to resolve outstanding questions about the origin of early kilonova emission \citep[e.g.][]{arcavi18}.  

Unlike previous LVC observing runs, O3 events are public and initial alerts are sent within minutes of discovery.  This allows any group to get immediate access to the GW candidate information, including classification probabilities (i.e. whether the event was a binary black hole merger, or contained a neutron star) and sky localization files.   In these first moments, dozens of research groups spring into action. 
This involves either tiling the localization region or targeting individual galaxies within it at the appropriate distance, depending on the resources available, and identifying new transients in real time.  There is very little coordination between groups which leads to both duplicated effort as well as large regions of the localization region that are likely inadequately searched.

While progress has been made to coordinate multi-telescope networks \citep[e.g.][]{coughlin19,grandma}, there is still a need for coordination across the entire electromagnetic follow-up community. The little coordination that does occur is through the Gamma-ray Coordinates Network (GCN)\footnote{\url{https://gcn.gsfc.nasa.gov/lvc.html}}  alert system, which usually consists of free-formed text that is very difficult to programmatically interpret. 

To solve this problem, we present the Gravitational Wave Treasure Map, an open source system for reporting, coordinating, visualizing, and assessing searches for, and the subsequent follow-up of, EM counterparts to gravitational wave events.  This system will reduce overlapping search efforts allowing the community to cover more sky area more efficiently. It will also automatically compute the total integrated probability searched. 

In Section~\ref{section:overview} we provide an overview of the Treasure Map and the goals of the project.  In Section ~\ref{section:components} we discuss the individual components of the Treasure Map.  In Section~\ref{section:future} we discuss the future functionality of the system, and we provide our conclusions in Section~\ref{sec:conc}. The success of this project relies on the engagement of the astronomical community partaking in these follow-up observations. The more users that participate in this system, the more it will flourish, allowing more efficient EM counterpart searches.

\section{Treasure Map Overview \& Goals}\label{section:overview}

\begin{figure*}[!b]
\centering
\includegraphics[width=6.5in]{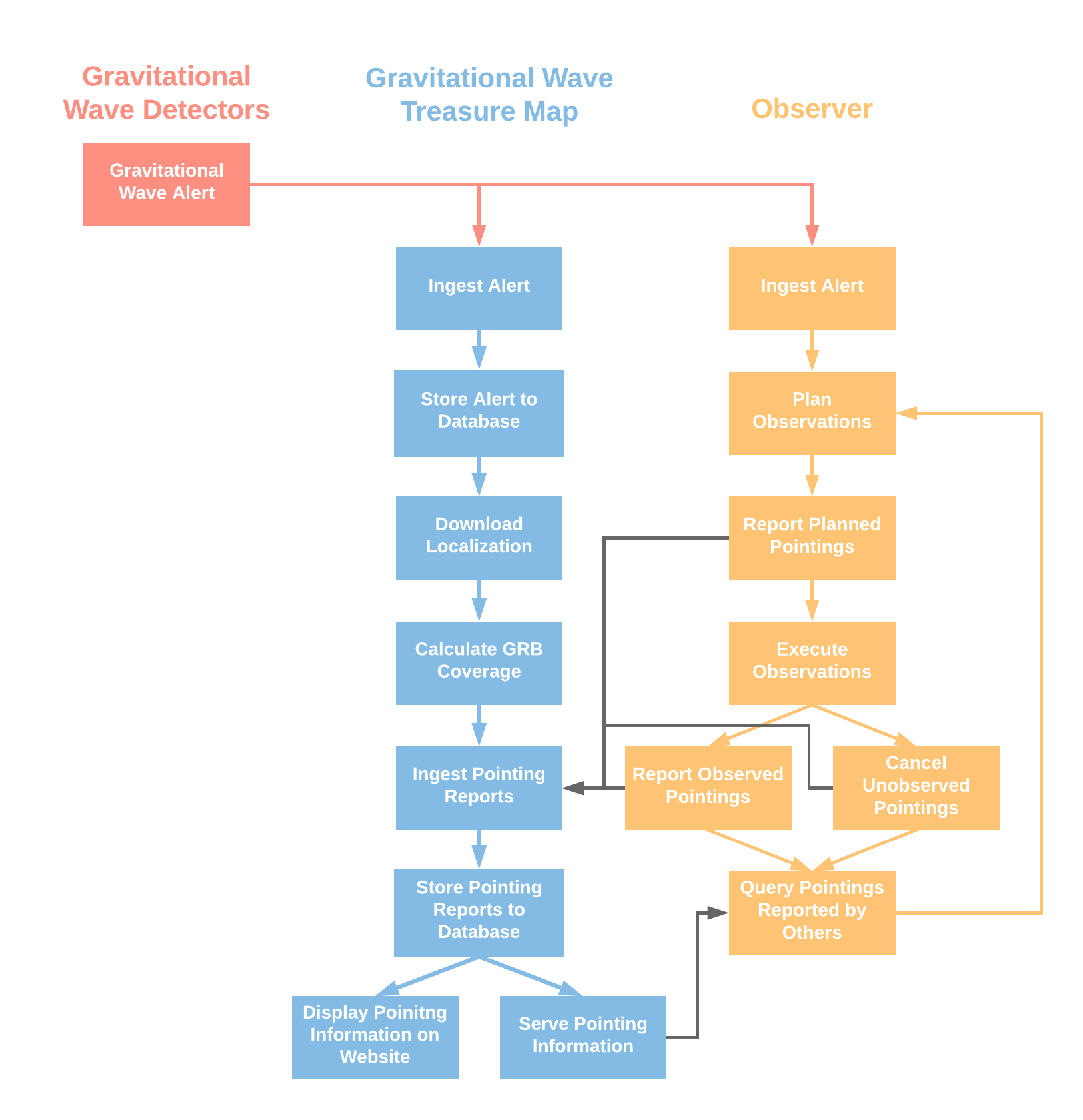}
\caption{Flowchart of the Gravitational Wave Treasure Map and its interaction with electromagnetic follow-up observers.  Users can report their planned and observed pointings with their specific instrument, as well as query the pointings posted by other groups, which can then feedback and inform new observations in neglected regions of the GW localization region. \label{fig:flow} }
\end{figure*}

The Treasure Map is a tool to coordinate, visualize, and assess the electromagnetic follow-up of gravitational wave events.  
To avoid users having to download and install software, and to ensure it is available across all computing platforms, the Treasure Map is available on the web at  \href{http://treasuremap.space}{http://treasuremap.space}. As it is meant to be a community tool, the code is open source\footnote{\url{https://github.com/swyatt7/gwtreasuremap}}.  Figure~\ref{fig:flow} uses a flow chart to illustrate the basic functionality. 
Here we give a brief overview of the Treasure Map before detailing its components in the next section.

Anyone can explore the Treasure Map web site and visualization service, however users must register an account to post their own pointings and query other observations via the API.  Users must also submit details about any counterpart search instrument they use (e.g. its name and footprint shape) if it does not already exist in the database. 

\begin{figure*}[!b]
\centering
\includegraphics[width=6.5in]{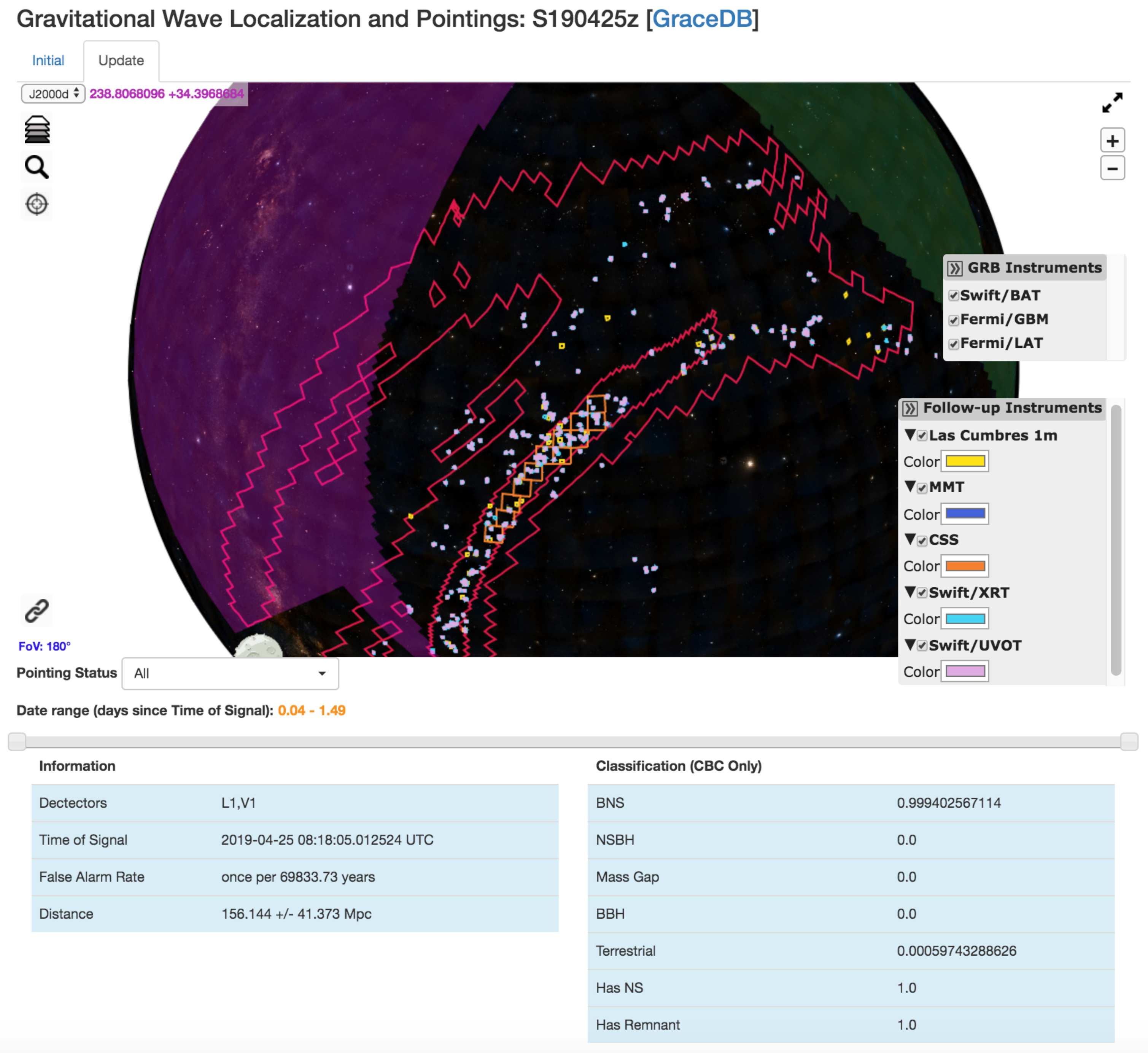}
\caption{Visualization for S190425z \citep{GCN24168}.  The LIGO/Virgo localization contours are shown in red (50\%, 90\%).  Other color outlines show telescope pointings which can be toggled on and off on a per-instrument basis.  Users may scroll and zoom and choose one of 24 backgrounds showing data from the X-ray to the IR (Mellinger survey base layer shown). Tabs (upper left) show all available localization refinements, e.g. {\em Initial} or {\em Update}. The purple and green shaded areas are the regions that \emph{Fermi/GBM} and \emph{Swift/BAT} were sensitive to at the time of the merger, respectively. The blue table shows relevant information about the event.  Just above the blue table is a time slider which can be used to visualize the sequence of observations taken to follow-up this event.  \label{fig:Vis} }
\vspace{-0.2in}
\end{figure*}

Once a gravitational wave alert has been issued by the LVC, it is ingested and stored in the Treasure Map database, along with the HEALPix map of the sky localization.  We also calculate the sky coverage of the {\it Fermi} Gamma-Ray Space Telescope's Gamma-ray Burst Monitor \citep[GBM;][]{GBM} and Large Area Telescope \citep[LAT; ][]{LAT} along with the Neil Gehrels {\it Swift} Observatory's Burst Alert Telescope \citep[BAT;][]{BAT} at the time of the gravitational wave signal.
For each event, a web-page is generated to visualize the localization region and the planned/completed observations of reporting groups (Figure~\ref{fig:Vis}).

As observers plan their EM counterpart searches, they can submit their telescope pointings to the Treasure Map via the API or website.  These planned pointings can be cancelled if they are never completed (i.e. bad weather, instrument problems, or simply a change in plans), and the set of final executed pointings can be uploaded.

Users can also request a digital object identifier (DOI) associated with their executed observations for a given event, allowing their observations to be cited by others.  Observers can query the Treasure Map database to determine the planned or executed observations of other groups, using this information to plan their own strategy.  These observations are visible on the website for each event with a double handled time slider and coverage calculator to show the community's unfolding observing campaign.

The Treasure Map was written during the first phase of the O3 run, and all O3 events have been ingested into the database, along with the pointed follow-up observations of several groups that helped with beta testing (e.g. the Berger Time Domain Group, \citealt{Hosseinzadeh19,Gomez19}; SAGUARO, \citealt{saguaro}; the Las Cumbres Observatory Gravitational Wave Follow-up team, \citealt{Arcavi17_2}; the \textit{Fermi} GBM and LAT teams; and the {\it Swift} GW follow-up team, \citealt{Klingler19}).  As of the start of O3b (1 Nov 2019), the Treasure Map has been `live' accepting planned and executed observations from the community \citep{GWTM_GCN}.

\section{Treasure Map Components}\label{section:components}

In order to facilitate cooperation and ultimately aid in identifying electromagnetic counterparts, the Treasure Map   is meant to be a real-time, central repository for all planned and completed observations for each GW event.
We next describe the components of this system.

\subsection{GW Alert Listener \& Gamma Ray Burst Mission Coverage}

When a candidate GW event is detected, the LVC sends out a VOEvent alert through the NASA Gamma-Ray Coordinates Network (GCN)\footnote{\url{https://gcn.gsfc.nasa.gov/lvc.html}}, along with a HEALPix localization map with distance constraints \citep{Singer16}.  The alert contents are described in the LIGO/Virgo Public Alerts User Guide.\footnote{\url{https://emfollow.docs.ligo.org/userguide/index.html}}  The Treasure Map ingests the LVC gravitational wave alerts in real time, along with their updates.  We also listen for alerts on a backup host in case the primary is down to ensure robustness.  The contents of each GW alert are displayed on a dedicated web-page for each event, along with a visualization of the 50th and 90th percentile sky contours, as seen in Figure~\ref{fig:Vis}.

 Short gamma-ray bursts were long suggested to be associated with neutron star mergers \citep[see, for instance, the short GRB review of][and references therein]{Berger14}, and this was confirmed by GW170817 \citep{GRB_GW170817}.  Thus, knowing the sky coverage of gamma ray telescopes at the time of a GW event is crucial knowledge which the Treasure Map makes available.
 The sky coverage for {\it Fermi}/GBM, \textit{Fermi}/LAT, and {\it Swift}/BAT is calculated and displayed automatically upon ingestion of a gravitational-wave alert. For the GBM instrument, this is done by calculating the instantaneous position of the {\it Fermi} spacecraft at the GW trigger time, using the most recent publicly accessible two-line element provided by CelesTrak\footnote{\url{https://www.celestrak.com/}}. The coverage area is then calculated as the all-sky area not below the Earth-limb with respect to the spacecraft at the trigger time. We similarly calculate the coverage of the {\it Swift}/BAT and \textit{Fermi}/LAT instruments based on where the satellites were pointing at the time of the event.

\subsection{API}\label{sec:api}

The Treasure Map API allows users to report their own observations and query the pointings from everyone participating in the counterpart search.
Once a user creates and verifies their account, they will be issued a unique token that will give them access to the API endpoints.  Currently, our API holds functionality for users to POST/GET pointings, UPDATE a pointing status to cancelled, GET instrument information, and POST a DOI for completed pointings (which is discussed in the following section). While the API is a major motivation for this project, we still offer all of its functionality via web forms, allowing users who do not want to submit and query pointings programmatically to still be able to use our service.

The API is well documented (see Section~\ref{sec:doc}), and there is a gravitational wave `test event' (labelled \code{TEST\_EVENT}) that serves as a code sandbox for incorporating  the Treasure Map into a group's workflow.  

\subsection{Database}\label{sec:database}

We use a PostgreSQL database to store our tables. We chose this over other databases to capitalize on its ability to perform spatial calculations within the queries themselves through the PostGIS library extension\footnote{\url{https://postgis.net/}}. When serializing the positions of each of our pointing objects, they are stored as a spatial POINT instead of individual float RA and DEC fields. This drastically reduces the computational time for SQL queries that rely on spatial calculations.

Figure \ref{fig:ERD} shows the Entity Relationship Diagram (ERD) for the Treasure Map database. The ERD shows the relationship between each entity as they appear as tables in the PostgresSQL database, along with each field and datatype in the tables. We chose to have a many-to-many relationship between the \code{gw\_alert} table and \code{pointing} tables since there could be  multiple gravitational wave alerts localized within the same sky region at around the same time. With this approach, a single pointing could  refer to multiple alerts. Also we designed our \code{instrument} table to have a many relationship to the \code{footprint\_ccd} table so that if an instrument has multiple CCDs defining its footprint, it can be visualized, and coverage calculations performed accordingly.

\subsection{The Web Application}

The Treasure Map web application is vital for user interactions and visualizing GW data.  The web site is built within a Python Flask WSGI\footnote{\url{https://pypi.org/project/Flask/}} web application framework, utilizing the  Flask SQLAlchemy object-relational mapper (ORM) to communicate to our database (described in Section~\ref{sec:database}).  Additionally, we use the functionality of the cloud based Amazon Web Services (AWS)\footnote{\url{https://aws.amazon.com/}} to host both our database (using Amazon's Relational Database Service, RDS) and web server (a spot request Elastic Compute Cloud, EC2, instance). The server is running an Ubuntu 18.04\footnote{\url{http://releases.ubuntu.com/bionic/}} instance and the website is served  through an Apache2 HTTP Server \footnote{\url{https://httpd.apache.org/}}. 

\begin{figure*}[!b]
\centering
\includegraphics[width=6.5in]{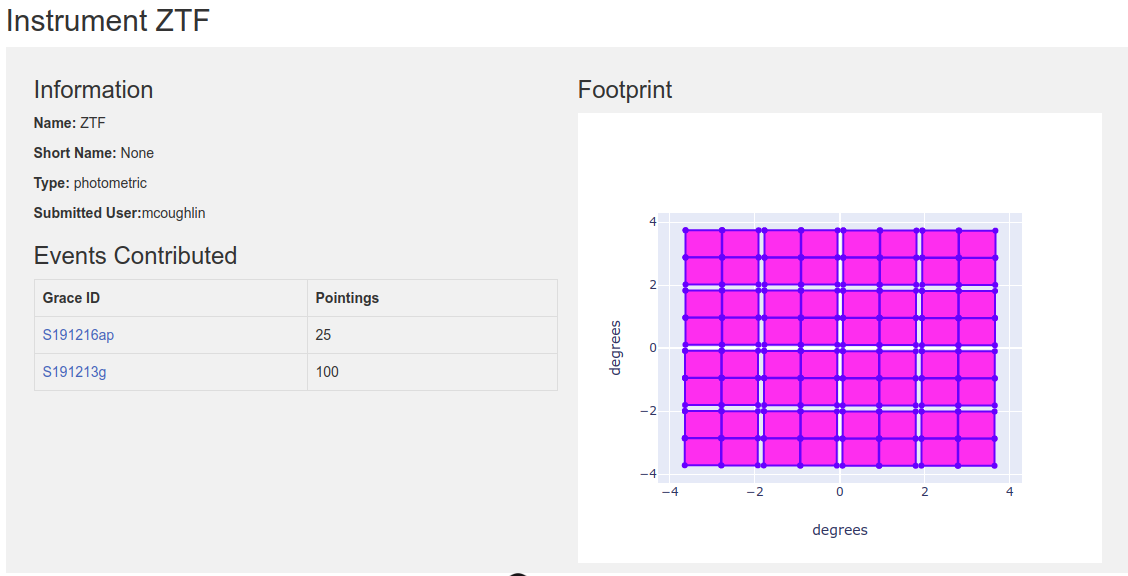}
\caption{ After registering, users may create instrument profiles (or use one that has already been created) to be associated with their planned/completed observations.  A simple web form allows the spatial parameters of the instrument footprint to be specified using circles, rectangles or more complicated polygons; chip gaps can also be represented. Above we show a screenshot of the ZTF camera, contributed to the Treasure Map site, with 64 individual CCD polygons. The chip gaps are accurately rendered onto the Aladin vizualization and taken into account for probability coverage calculations. \label{fig:Inst} }
\vspace{-0.2in}
\end{figure*}

The front-end of the website allows for a user to: create an account, post/query instruments, post/query observations, request DOIs for completed observations, and visualize all available aspects of each ingested gravitational wave alert. Once a user creates an account, they are issued an API token upon verification. Having a verified account allows the user to submit their instrument to the Treasure Map database which can be referenced upon each submission of a planned/completed observation. There are three methods for constructing instrument footprints:  rectangular, circular, or a multi-order polygon.  This simplifies the process of creating complex footprints such as for multi-CCD mosaic instruments. We show an example of a submitted instrument in Figure~\ref{fig:Inst}.  The footprint plays a crucial role in the visualization of the observations and the coverage calculations (Section~\ref{sec:calc}).

\begin{figure*}
 \centering
 {
 \parbox{3.2in}{\centering\includegraphics[width=3.2in]{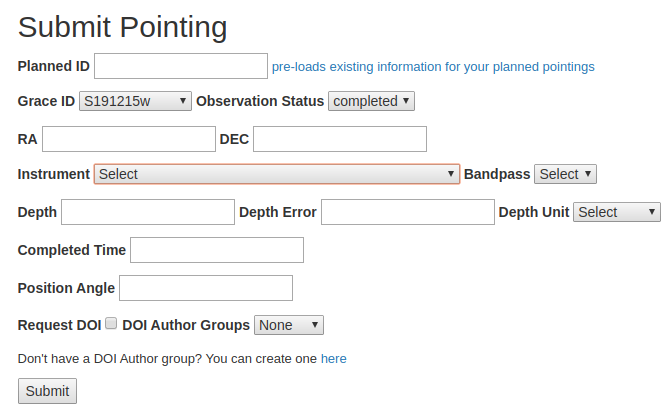}}
 \parbox{3.2in}{\centering\includegraphics[width=3.2in]{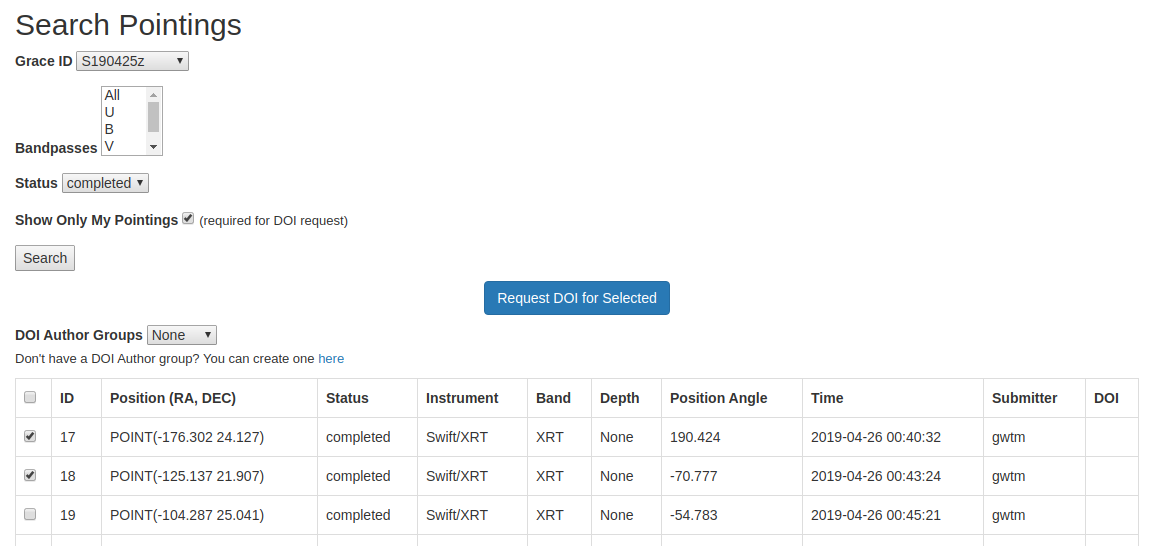}}
 }
 \caption{We provide a graphical user interface for submitting and downloading telescope pointings, although we expect most users to access this programmatically via the API. {\em Left:} The pointing submission GUI. {\em Right:} Searchable pointing information with the option to batch request a DOI for completed pointings. \label{fig:SubmitPointing} }
 \end{figure*}

Once a user has their instrument(s) saved, they can post their pointings either through the `Submit Pointing' web-form or the programmatic API POST method (described in Section~\ref{sec:api}). Each of these methods are documented on the website (see Section~\ref{sec:doc}). Users also have access to the `Search Pointings' web-form which displays a table of pointing observation information (including the position on sky, status, instrument, band, depth, position angle, time, submitter and DOI reference) for each requested gravitational wave event. The same functionality is available through the API GET method which returns a JSON list of pointing objects. We also allow users to query for all available instrument information. This allows users to perform their own visualization of an ongoing search, perform their own spatial calculations, or even optimize their own observational scheduling.  Each of these web-form functions are shown in Figure \ref{fig:SubmitPointing}.

\subsection{Citable Digital Object Identifier}

We have also added the ability to cite completed observations through the Digital Object Identifier (DOI) service from the open science platform Zenodo\footnote{\url{https://about.zenodo.org}} so that groups can get credit for their work when follow-up observations are discussed in the literature. A DOI may be requested when a group submits their pointing(s) through the API or through the 'Submit Pointings' webform. We also have the capability to request a DOI after submission through the API \code{request\_doi} endpoint or via `Search Pointings' web-page as seen in Figure \ref{fig:SubmitPointing}. When uploading the DOI, the object is serialized as an open-access dataset that is represented as a \code{json} file analogous to the API GET request for the same pointings. Before requesting the DOI, the user has the ability to create a list of authors for the citable dataset or submit an author list through the Treasure Map API POST request. Once the DOI is successfully submitted, the user is returned a URL that will give them access to the citation information.  Several groups have requested DOIs thus far in O3 \citep{iair_arcavi_dataset,michael_lundquist_dataset,aaron_tohuvavohu_dataset}.  

\begin{figure*}[!b]
\centering
\includegraphics[width=6.3in]{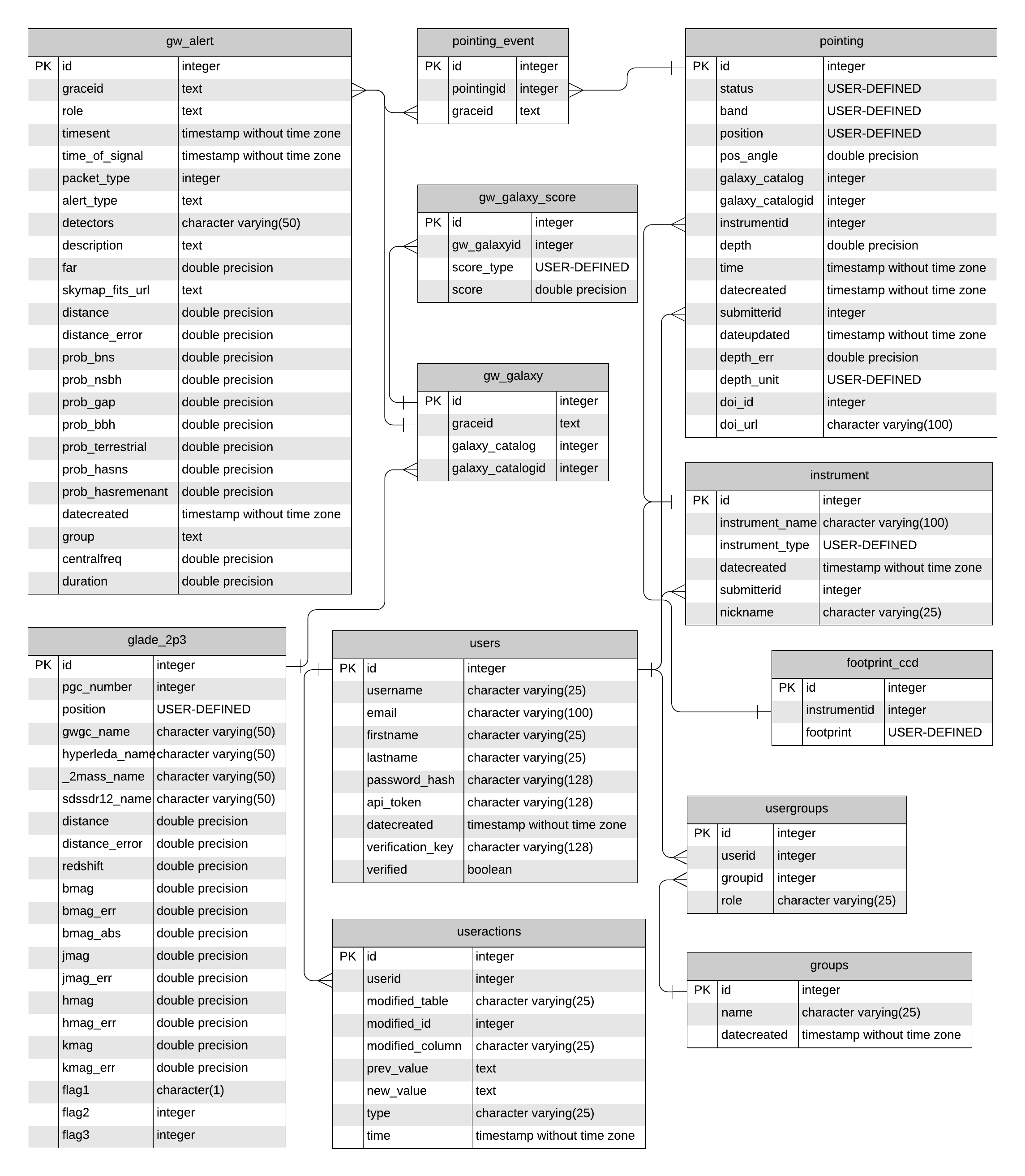}
\caption{Entity Relationship Diagram for the Treasure Map PostgresSQL database, showing each  field and datatype in the tables, along with the relationship between the tables. The connections between tables indicate their relationship, for instance the three-pronged symbol is a 'many' relationship, and the perpendicular hash mark indicates a 'one' relationship.  A one-to-many relationship would indicate that a row in the first table may be linked to many rows in the second table, while a many-to-many relationship may have multiple rows in each table linked to each other.  We have tables for galaxy association with GW events (e.g. the \code{glade\_2p3}, \code{gw\_galaxy} and \code{gw\_galaxy\_score} tables), which are not fully implemented yet, but will be in a future iteration of the Treasure Map (Section~\ref{section:future}).  \label{fig:ERD} }
\vspace{-0.2in}
\end{figure*}

\begin{figure*}[!b]
\centering
\includegraphics[width=6.5in]{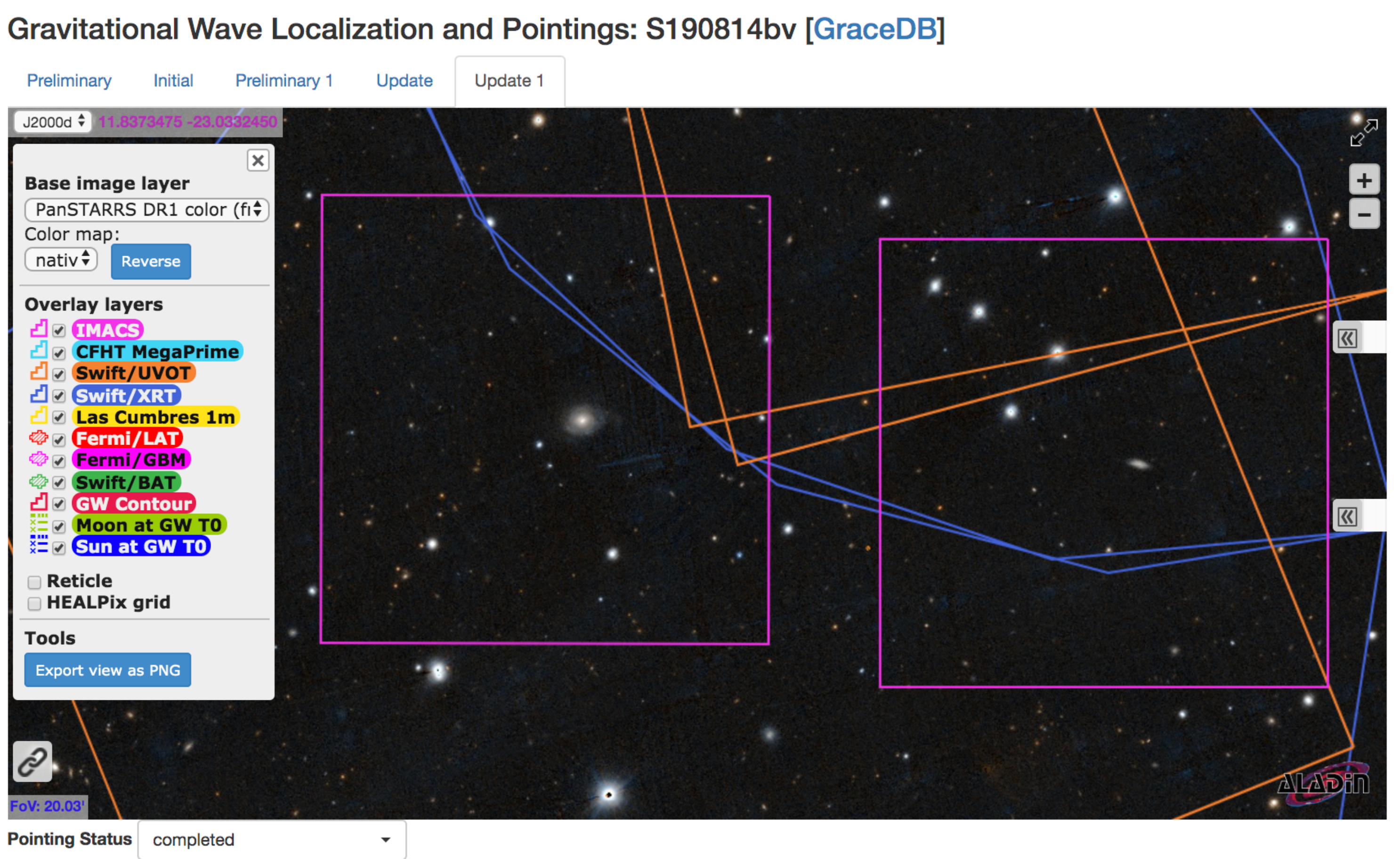}
\caption{  Zoomed in Treasure Map visualization for S190814bv \citep{gcn25324} with Pan-STARRS DR1 as the base layer -- users may choose from 24 different image survey background layers, spanning the gamma-ray to infrared regimes.  Users may select which pointings to show, the base layer, and the sun and moon position at the time of the trigger using the menu on the left. The base image layer functionality available through Aladin Lite can provide real time information to users, for instance displaying the positions of known X-ray sources within the GW localization. \label{fig:Vis2} }
\vspace{-0.2in}
\end{figure*}

\subsection{Visualization}\label{sec:vis}

We use Aladin Lite \citep{Alite} for interactive visualization of gravitational wave localizations and pointing information on the Treasure Map web site.  We show two different illustrations of our visualization in Figures \ref{fig:Vis} and \ref{fig:Vis2}, displaying different background imaging surveys for each. We chose this application because it can efficiently overlay LVC skymap HEALPix contours, instrument footprints, and multi-order coverage (MOC) maps upon imaging survey data in a two dimensional spherical projection.  It is also widely used in the astronomical community.  The application has an easy-to-use Javascript API and is powered by the HTML5 canvas technology which is compatible with any modern browser. It uses standardized Hierarchical Progressive Survey (HiPS) technology \citep{Fernique15}, which supports tiling and zooming, loading progressively higher resolution data as needed. 

It also supports HEALPix projections \citep{healpix}, used by the LVC to distribute localizations.

The Aladin Lite visualization tool is very customizable, allowing users to view the localization in many different ways including panning across the projection and zooming into the map. There are twenty-four image survey base layers available directly from the Aladin Lite service, including standard optical and infrared surveys, H$\alpha$ maps, and high energy X-ray and gamma-ray surveys; these may provide additional contextual information to users during their search. Figure~\ref{fig:Vis2} shows a zoomed Treasure Map view with a Pan-STARRS DR1 \citep{pstars1} base layer.  We allow users to toggle any overlay that is loaded onto the map, including the localization region contours, instrument footprints, and the GRB mission coverage MOC maps. Along with toggling the instrument footprints, users have the ability to change the loaded outline colors of the instrument footprints. Both planned and completed pointing data can be visualized.  There is also a time slider which dynamically loads the instrument footprints giving the user a real-time playback of when observations occurred.

\subsection{Coverage Calculator}\label{sec:calc}

\begin{figure*}[h]
\centering
\vspace{0.3in}
\includegraphics[width=6.5in]{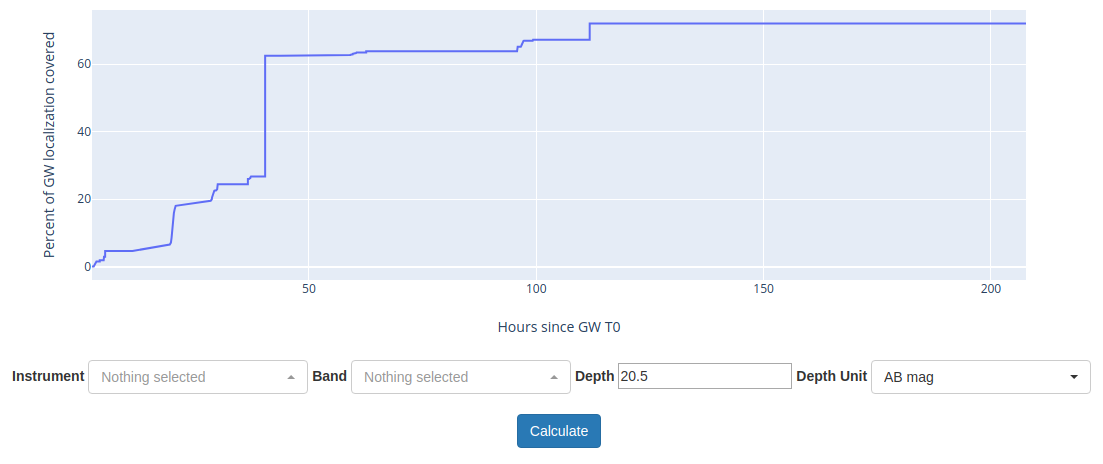}
\caption{Coverage calculation of the very well localized event S190814bv. This event had a localization region of around 23 deg$^{2}$ and had around 1800 reported completed pointings into the Treasure Map database from 5 different instruments. The plot shows the percentage of the localization region covered versus time (in hours since time of signal). The user can filter out pointings based on the parameters shown above. Here we show an example of the coverage for this event with a limiting depth of 20.5 AB magnitude. \label{fig:CovCalc} }
\vspace{-0.2in}
\end{figure*}

For each gravitational wave event, we provide a coverage calculator so that users can plot the probability percentage of the localization region that is covered as a function of time after the GW detection. This can be used to assess how successful early coverage of the localization region was and also see what probability was observed to a certain depth. The tool is located beneath the visualization on the individual alerts page, and has several customizable parameters.  Users can filter based on specific instruments, instrument filters, and search depth. An example of this tool is shown in Figure~\ref{fig:CovCalc} for S190814bv showing that within 48 hours of the detection of S190814bv, $\approx 60$\% of the localization region had been observed by telescopes reporting to the Treasure Map with a limiting depth of 20.5 AB magnitude. 

The coverage calculator works by iterating over each completed pointing's instrument footprint and querying where it lies in the HEALpix map by using the healpy function \code{query\_polygon}. This function returns the pixels whose centers lie within the convex polygon defined by the footprint's vertices array.\footnote{Note that this is an inherently conservative approach. There will be HEALpix pixels which are partially covered by an instrument footprint, but whose center does not lie within the coverage of the footprint. The probability contained in these pixels is not counted.} We test for duplicated pixel centers (to avoid double counting probability if the same area is covered by multiple instruments, or overlapping fields-of-view), and then iterate over them summing up the posterior probability associated with each covered pixel. We use the associated completed observation time as a timestamp.

\begin{figure*}[h]
\centering
\includegraphics[width=5in]{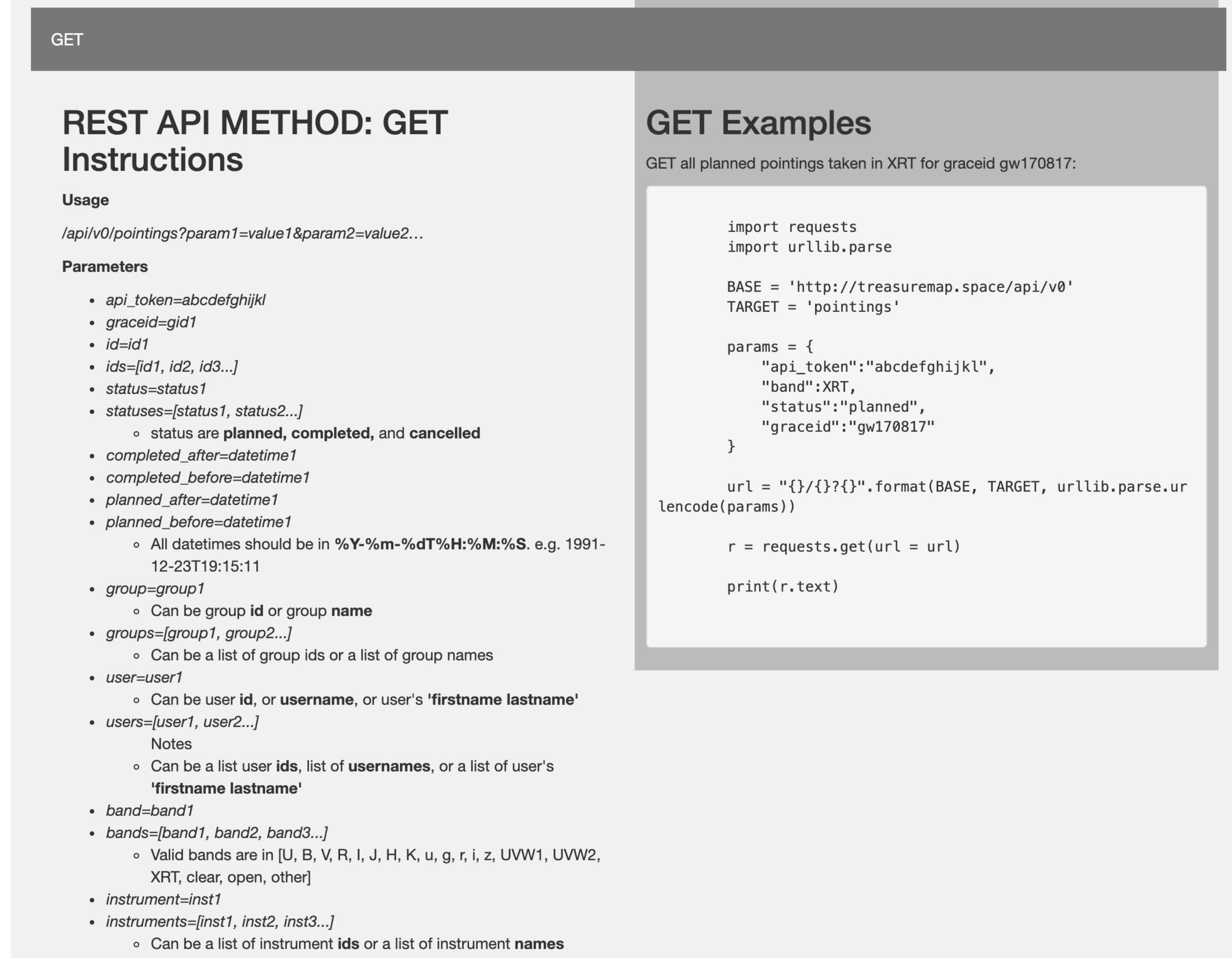}
\caption{The web site provides documentation for each API command, including example python code. Here is an example of the REST API GET Method for the pointings table.  This method is used to query the pointings taken of a given gravitational wave event, with the ability to filter on instrument, instrument band, planned/completion status, planned/completed time, pointing ID and user group.  \label{fig:Doc-Get}}
\end{figure*}

\subsection{Documentation}\label{sec:doc}

Our goal is to make the Treasure Map easy to use, with a small barrier to entry, enabling as many groups as possible to participate in this collaboration tool.  Documentation and how-to tutorials are thus a priority.  To aid this, each of our API endpoints are documented on the website (see, for example, Figure~\ref{fig:Doc-Get}).  Along with the documentation, example python code is also provided so that the users can have a working example of how to access each endpoint. Example jupyter notebooks are also available on the Treasure Map github repository to walk through example use cases. While our examples are all in python, any programming language that allows HTTP requests will be sufficient.  Finally, a ``What's New'' development blog\footnote{\href{http://treasuremap.space/development_blog}{http://treasuremap.space/development\_blog}} on the web site informs users of changes and upgrades to the Treasure Map's functionality.

\section{Future Functionality}\label{section:future}
There are several features that we intend to roll out as part of a Phase II for the Treasure Map in preparation for O4 and beyond.  These features take into account recommendations and feedback from users and the community.  
An initial list includes: 

\begin{itemize}

\item The Treasure Map plans to have all counterpart candidates that have been reported to the Transient Name Server\footnote{https://wis-tns.weizmann.ac.il/} ingested during an event. These objects will be overlayed on an events visualization page and their classification status will be represented with different symbols. 

\item Prioritized galaxy lists are crucial for narrow field EM counterpart searches, especially for nearby events \citep[e.g.][]{Gehrels16,Arcavi17,Yang19,Ducoin19}. When an alert is ingested, we will locate each galaxy that is within the GW distance estimate for each HEALpix pixel and assign a score to each cataloged galaxy therein (possibly using multiple scoring algorithms); a similar service is already available as a web tool \citep{hogwarts}. These galaxies will be available as an API endpoint allowing users to quickly retrieve galaxies and their scores. This relation is seen in the \code{gw\_galaxy} table in the ERD (Figure~\ref{fig:ERD}), where each galaxy located within the region will be assigned a score based on parameters defined for each entry in the \code{gw\_galaxy\_score} type table.  We already have the  Glade\_2p3 \citep{glade} galaxy catalog ingested into our database and will incorporate other galaxy catalogs in the future.  

\item The coverage calculator functionality will also be expanded to include the ability to calculate coverage of the galaxy-convolved localization regions, as an addition to the current capability of calculating coverage of the raw GW localization.

\item We also plan on cross-matching galaxy lists with already completed pointings.  This will provide users with lists of galaxies that have yet to be observed, available as an API endpoint.  Users will be able to sort by the score that our software provides or use other software algorithms to create their own prioritized lists.

\item We intend to ingest alerts and localizations from other relevant detectors, in particular from gamma-ray and neutrino observatories.  For instance, the O3 event \href{http://treasuremap.space/alerts?graceids=S191216ap}{S191216ap} had both a \textit{possible} coincident IceCube neutrino detection \citep{icecube_gcn} and subthreshold gamma-ray detection from the High-Altitude Water Cherenkov (HAWC) gamma-ray observatory \citep{hawc_gcn} with (nearly) overlapping localizations on the sky. There was also a coincident Fermi-GBM GRB detection roughly 10 minutes after the alert of the O3 event \href{http://treasuremap.space/alerts?graceids=S200219ac}{S200219ac} \citep{GCN27144}; the Treasure Map is currently hosting localization information on this event on request from the Fermi-GBM team. Visualizations of all relevant localizations at once, along with their convolved footprints, will be provided in a future iteration of the Treasure Map. 

\item We plan to develop TOM Toolkit \citep{tom_toolkit} support for the Treasure Map. TOMs are Telescope and Object Managers -- databases connected to observational data, visualizations and/or telescopes.  They allow you to plan and execute new observations, and automatically reduce and visualize incoming data. The TOM Toolkit is an open source set of tools for creating TOMs\footnote{https://lco.global/tomtoolkit/}. TOM Toolkit support will allow users to deploy their own Treasure Map visualization native to their existing TOM. This would allow them to, for example, plot proprietary data, and possibly even visualize their own reduced data (not just the footprint) projected on the visualization to see it in context.

\item We plan to incorporate the functionality of alerting users to various subscribable events. The alerts could be from GW event updates like a counterpart being discovered, updated localization regions, or new services that are not yet incorporated.

\item We plan to provide a service that returns the best region to observe given what has already been observed and your instrument parameters.  This may be a galaxy that has not yet been observed or assigned, or the highest unobserved probability region of the localization.

\item We are examining the addition of the Virtual Observatory's Observation Locator Table Access Protocol (ObsLocTAP) into the Treasure Map, as this provides a data model and access protocol for communicating metadata about astronomical observations through the widely-accepted Virtual Observatory framework.


\end{itemize}
\section{Conclusion}\label{sec:conc}

In this paper we have described the Gravitational Wave Treasure Map, a tool for coordination and visualization of EM counterpart searches to gravitational wave events.  The Treasure Map  provides a single resource for astronomers to both share and query observational pointings in their search for EM counterparts. 

With a responsive visualization tool, comprehensive API, and citable DOIs for completed pointings, the Treasure Map is an extensive package for GW EM counterpart searches.

Expansion plans for the Treasure Map include the ability to share and query individual electromagnetic counterpart candidates and their followup to further improve coordination and to avoid duplicate observations.  We will also provide catalog lists and visual representations of individual galaxies coincident with GW localization regions.  Future implementations of the Treasure Map will include constraints and localizations from other multi-messenger relevant facilities (e.g. neutrino and gamma-ray observatories) that may contribute to the hunt for GW counterparts.  We are also open to community feedback and will consider implementing suggested features.

Some cultural change in the multi-messenger astronomy community may be necessary to fully incorporate collaboration tools such as the Treasure Map.  However, results from O3 have been heartening, with large scale participation in the GCN system by the community.  Additionally, many major observing collaborations are already participating in the Treasure Map, which can be seen as an extension of what is available via GCNs.  There is thus reason for optimism that the field will continue in a collaborative direction.


The Treasure Map is just one part of a larger cyberinfrastructure ecosystem that should be built to facilitate multi-messenger astrophysics \citep[e.g.][]{MMA_1,MMA_2}, and several groups have formed to plan these endeavors (e.g.~the Scalable Cyberinfrastructure to support Multi-Messenger Astrophysics collaboration; SCiMMA\footnote{\url{https://scimma.org}} and the planned GCN upgrade Time Domain Astronomy
Coordination Hub; TACH).  These efforts are essential to make the most of the current multi-messenger era.

\vspace{0.25cm}

\acknowledgments

We want to acknowledge Curtis McCully and Austin Riba who both assisted and provided insight in the early development and management of the web server.  We also want to acknowledge Joseph Long who assisted with projection plotting near the poles of the Aladin Lite visualization, and debugging an Apache2 multi-thread deadlock issue. We also acknowledge helpful discussion and the contribution of code from Daniel Kocevski, which was critical to the implementation of automatic calculation of \textit{Fermi} GBM and LAT coverage. Further useful discussions with the SCIMMA team, as well as J. Brown, D. Coulter, and R. Foley helped the development of the Treasure Map.

This work was supported in part by the Israel Science Foundation under grant No. 2752/19). IA is a CIFAR Azrieli Global Scholar in the Gravity and the Extreme Universe Program and acknowledges support from that program, from the Israel Science Foundation (under grant No. 2108/18 and 2752/19), from the United States - Israel Binational Science Foundation (BSF), and from the Israeli Council for Higher Education Alon Fellowship. Research by DJS is supported by Natinal Science Foundation grants AST-1821967, 1821987, 1813708, 1813466, and 1908972.  Discussions were held at the Kavli Institute for Theoretical Physics, which is supported under NSF grant No. PHY-1748958


\software{Astropy \citep{astropy}, Ephem, Flask, Healpy \citep{healpy}, \code{ligo.skymap} \citep{Singer16},  NumPy \citep{numpy}, Plotly \citep{plotly}, Python \citep{python},  PyGCN, SQLAlchemy \citep{sqlalchemy}}

\bibliographystyle{aasjournal}
\bibliography{gwtm}

\end{document}